\documentclass[a4paper,11pt]{article}
\usepackage{jcappub}
\usepackage{bm}
\usepackage{amsthm,xcolor}

\theoremstyle{definition}
\newtheorem{example}{Example}[section]
\newtheorem{remark}{Remark}[section]

\DeclareMathOperator*{\Laplace}{Laplace}
\providecommand*{\iu}%
{\ensuremath{\mathrm{i}}}
\def\diff{\mathrm{d}}

\title{\boldmath Kinetic field theory of compact systems}

\author[1]{Matthias Bartelmann}
\affiliation{Institute for Theoretical Physics, Heidelberg University, Philosophenweg 12 and 16, 69120 Heidelberg, Germany}

\author[2]{James Stokes}
\affiliation{Department of Mathematics, University of Michigan, Ann Arbor, MI, United States of America}

\abstract{The kinetic field theory is developed without assumptions of statistical homogeneity and isotropy. In a solvable toy model with short-ranged interactions, we compare first-order perturbation theory to an iterated mean-field approximation scheme, demonstrating that the mean-field theory maintains positivity and captures collapse dynamics, allowing analytic estimates of blow-up times. In a self-gravitating sheet model, the first-order perturbation theory is shown to reproduce critical phenomena. This work suggests a path toward convergence analysis of the mean-field approximation and applications to more complex inhomogeneous systems.}

\begin{document}
\maketitle
\flushbottom

\newpage
\section{Introduction}

The statistical mechanics of many‐body systems far from equilibrium poses a fundamental challenge across a wide range of disciplines, including plasma physics, cosmology and condensed matter. In recent years, Kinetic Field Theory (KFT) \cite{mazenko2010fundamental, das2013newtonian, bartelmann2017kinetic, fabis2018kinetic, bartelmann2019cosmic, kozlikin2021first, bartelmann2021kinetic, konrad2022asymptotic} has emerged as a powerful framework for describing the evolution of classical particle ensembles via a generating functional for microscopic trajectories.

Most applications of KFT have focused on cases in which the initial phase-space density is statistically homogeneous and isotropic; a standard assumption in cosmic structure formation (CSF) \cite{bartelmann2016microscopic} as well as in plasma physics applications \cite{kozlikin2023ultracold}. In these studies, the initial phase-space density includes inverse volume factors that pair up with the particle number $N$ to yield the density $N/V$, which remains finite as $N\to\infty$. After marginalizing over momentum space, the initial phase-space density reduces to a uniform distribution.

In this work, we extend KFT by lifting the assumption of statistical homogeneity and isotropy, making the framework applicable to isolated stellar systems or `island universe' cosmologies. Rather than pursue realistic astrophysical models, we employ analytically tractable toy models to showcase the purely theoretical innovations of the formalism. In particular, both first-order perturbation theory (FOPT) and an iterated mean-field approximation (MFT) are developed. In a solvable toy model with short-ranged interactions, we compare their regimes of validity and argue that MFT accurately captures the late-time dynamics. As a further case study, we examine a self-gravitating sheet model, showing that MFT breaks down on the collapse timescale while FOPT successfully reproduces critical phenomena.

The paper is organized as follows.  In section~\ref{sec:theory} we review the KFT framework and approximation schemes based on first-order perturbation theory and mean-field theory.  Section~\ref{sec:examples} presents several simple examples, including short-ranged interactions and self‐gravitating sheets. Section \ref{sec:discussion} discusses our conclusions. Many technical details such as perturbative calculations are collected in the appendices.

\section{Theory}\label{sec:theory}
This section summarizes the mathematical formulation of kinetic field theory and develops approximation schemes based on first-order perturbation theory and mean-field theory, which are suitable for compact systems.

In its most general form, the kinetic field theory concerns the statistical mechanics of an ensemble of $N \gg 1$ identical point particles undergoing Hamiltonian dynamics on a Riemannian manifold $M$. In this section we take $M=\mathbb{R}^d$,
so that the phase space admits coordinates of the form $\bm{x}=(x_1,\ldots,x_N) \in \mathbb{R}^{2d} \otimes \mathbb{R}^N$ with $x_j =(q_j,p_j)\in\mathbb{R}^{2d}$. The initial state of the system at time $t=0$ is a probability density function $\rho_0$ on phase space, which evolves under Liouville (Hamiltonian) dynamics. The identical nature of the particles is reflected in the following exchangeability assumption for $\rho_0(\bm{x})$ and the Hamiltonian $H(\bm{x},t)$,
\begin{equation}\label{e:exchangeable}
    \forall\pi \in S_N :
    \left\{
    \begin{aligned}
        \rho_0(\bm{x}^\pi) & = \rho_0(\bm{x}) \\
        H(\bm{x}^\pi,t) & = H\big(\bm{x},t\big)
    \end{aligned}
    \right.,
\end{equation}
where $\bm{x}^\pi = \big(x_{\pi(1)},\ldots,x_{\pi(N)}\big)$. The exchangeability assumption \eqref{e:exchangeable} implies that the solution of the Liouville equation is exchangeable for all $t>0$.
The time-dependent Hamiltonian is moreover assumed to admit an additive decomposition $H = H_0 + H_1$, where $H_0$ is an exactly solvable Hamiltonian, which serves as the expansion point for perturbation theory. Then the solution of the initial value problem
\begin{equation}
    \left\{
    \begin{aligned}
    \dot{\bm{x}}(t) & = \mathcal{J} \nabla H(\bm{x}(t),t) \\
    \bm{x}(0) & = \bm{x}_0
    \end{aligned}
    \right.
\end{equation}
satisfies the following recursive identity
\begin{equation}\label{e:recursion}
    \bm{x}(t) = \bm{G}(t,0)\bm{x}_0 + \int_0^t\diff u \, \bm{G}(t,u) \bm{F}(\bm{x}(u),u),
\end{equation}
where $\bm{F} = \mathcal{J} \nabla H_1$ and $\bm{G}$ is the Green function for $H_0$.

The quantity of primary interest in KFT is the so-called $n$-particle density correlator, which is defined as follows. For each $j \in [N]:=\{1,\ldots,N\}$, let $\tilde{\rho}_j(k,t) := e^{-\iu k \cdot q_j(t)}$ denote the Fourier transform of the density function $\rho_j(q,t) = \delta_\textrm{D}(q-q_j(t))$, which tracks the position of the $j$th particle.  Then the $n$-particle density correlator is defined as the expectation value with respect to $\bm{x}_0 \sim \rho_0$ of the Fourier-transformed densities for a subset of $n\leq N$ particles. By the exchangeability assumption \eqref{e:exchangeable}, the subset of particles can be chosen to be $[n]\subseteq[N]$, which gives rise to the expression
\begin{equation}\label{e:density_correlator}
    \underset{\bm{x}_0\sim\rho_0}{\mathbb{E}}\big[\tilde{\rho}_1(k_1,t_1)\cdots \tilde{\rho}_n(k_n,t_n)\big].
\end{equation}
Henceforth, we drop the subscript on the expectation value. Of particular interest is the equal-time density-density correlator $P(k,t)$, defined by setting $n=2$, $t_1=t_2 = t$ and $k_1 = k = -k_2$,
\begin{equation}\label{e:P(k,t)}
    P(k,t)
    := \mathbb{E}[\tilde\rho_1(k,t)\tilde\rho_2(-k,t)].
\end{equation}
In a statistically homogeneous setting, the restriction to diagonal wave vectors ($k_1 + k_2 = 0$) can be made without loss of generality because $\mathbb{E}\big[\rho_1(q_1,t)\rho_2(q_2,t)\big]$ only depends on the relative coordinate $q_1-q_2$. Although $P(k,t)$ does not capture all of the two-point structure in an inhomogeneous setting, it is nevertheless interesting to study because its inverse Fourier transform $P(q,t)$ retains a clear configuration-space interpretation. Specifically, $P(q,t)$ can be interpreted as the probability of finding $q_1(t)$ and $q_2(t)$ at separation $q\in\mathbb{R}^d$,
\begin{equation}\label{e:P(q,t)}
    P(q,t):=\int_k e^{\iu k \cdot q} P(k,t) = \int_k \mathbb{E}\big[ e^{\iu k\cdot(q_2(t) - q_1(t) + q)} \big] = \mathbb{E}\big[ \delta_{\rm D}\big(q_2(t)-q_1(t) + q \big) \big].
\end{equation}
For future reference, we also define the inertial correlator
\begin{align}
    P_0(k,t)
    := \mathbb{E}\left[e^{- \iu k \cdot ( \bar{q}_1(t) - \bar{q}_2 (t) )}\right],
\end{align}
which is defined in terms of the inertial trajectories $\bar{\bm{x}}(t) := \bm{G}(t,0)\bm{x}_0$.

In developing approximations for the $n$-particle density correlator \eqref{e:density_correlator}, it is useful to introduce a characteristic functional, defined as the following functional of a source field $\bm{J}(t) \in \mathbb{R}^{2d} \otimes \mathbb{R}^N$,
\begin{equation}\label{e:characteristic}
    Z[\bm{J}] := \mathbb{E} \exp\left[\iu \int_0^T\diff t' \bm{J}(t') \cdot \bm{x}(t') \right],
\end{equation}
where $T > t_j$ for all $j\in[n]$.
Then \eqref{e:density_correlator} can be expressed as
\begin{equation}
    \mathbb{E}[\tilde\rho_1(k_1,t_1)\cdots \tilde\rho_n(k_n,t_n)]
    = \hat \rho_1(k_1,t_1)\cdots \hat \rho_n(k_n,t_n) Z[\bm{J}]\Big|_{\bm{J}=0},
\end{equation}
where we have defined
\begin{equation}
    \hat{\rho}_j(k,t) := \exp\left[-\iu k \cdot \frac{1}{\iu}\frac{\delta}{\delta J_{q_j}(t)}\right].
\end{equation}
It is convenient to define a generating functional
$Z[\bm{J},\bm{K}]$ of two source fields $\bm{J}(t),\bm{K}(t) \in \mathbb{R}^{2d} \otimes \mathbb{R}^N$,
\begin{align}
    Z[\bm{J},\bm{K}] & := \mathbb{E} \exp\left[\iu \int_0^T\diff t' \bm{J}(t') \cdot \bar{\bm{x}}[\bm{K}](t') \right], \\
    \bar{\bm{x}}[\bm{K}](t) & := \bm{G}(t,0) \bm{x}_0 + \int_0^t \diff u \, \bm{G}(t,u) \bm{K}(u). \label{e:xbar}
\end{align}
It follows from the recursive property \eqref{e:recursion} that the characteristic functional can be expressed in terms of the generating functional as follows,
\begin{align}
    Z[\bm{J}]
    & = \left.\exp\left[\int_0^T \diff t' \bm{F}\big(\bar{\bm{x}}[\bm{K}](t'),t'\big) \cdot \frac{\delta}{\delta \bm{K}(t')}\right] Z[\bm{J},\bm{K}]\right|_{\bm{K}=0}, \\
    & = \left.\exp\left[\int_0^T \diff t'\bm{F}\left(\frac{\delta}{\iu \delta \bm{J}(t')},t'\right) \cdot \frac{\delta}{\delta \bm{K}(t')}\right] Z[\bm{J},\bm{K}]\right|_{\bm{K}=0},
\end{align}
and thus
\begin{align}\label{e:n_particle}
    \mathbb{E}[\tilde\rho_1(k_1,t_1)\cdots \tilde\rho_n(k_n,t_n)]
    & = \left. \hat \rho_1(k_1,t_1)\cdots \hat \rho_n(k_n,t_n) \exp\left[\int_0^T \diff t'\bm{F}\left(\frac{\delta}{\iu \delta \bm{J}(t')},t'\right) \cdot \frac{\delta}{\delta \bm{K}(t')}\right] Z[\bm{J},\bm{K}]\right|_{\bm{J},\bm{K}=0}.
\end{align}
\subsection{Approximation schemes}
In order to make progress in approximating \eqref{e:density_correlator}, we now impose additional structure on the Hamiltonian.
In particular, we choose the exactly solvable Hamiltonian to be a diagonal quadratic form in the phase space coordinates,
\begin{equation}
    H_0(\bm{x},t) = \frac{1}{2}\sum_{i=1}^N x_i\cdot h(t) x_i.
\end{equation}
The Green function then evaluates to
\begin{equation}\label{e:Green}
    \bm{G}(t,t') = G(t,t') \otimes I_N,
\end{equation}
where the single-particle Green function is given by
\begin{equation}
    G(t,t')=\mathsf{T}\exp\left[J\int_{t'}^t \diff u \, h(u)\right],
\end{equation}
and where $\mathsf{T}$ denotes the time-ordered exponential. In addition, we choose a momentum-independent interaction Hamiltonian of two-body form,
\begin{equation}
    H_1(\bm{x},t) = \frac{1}{2N}\sum_{i\neq j = 1}^N v\big(q_i-q_j,t\big),
\end{equation}
where $v$ is a parity-invariant interparticle potential
\begin{equation}\label{e:parity}
    v(-q,t) = v(q,t).
\end{equation}
The $1/N$ prefactor in the potential is necessary to define the large-$N$ limit. It may be helpful to consider the special case of the gravitational $N$-body problem, expressed in terms of velocity variables,
\begin{equation}
    H = \frac{1}{2}\sum_{i=1}^N m v_i^2 - \frac{1}{2}\sum_{i \neq j = 1}^N\frac{Gm^2}{|q_i-q_j|},
\end{equation}
where phase space coordinates are now $x_j = (q_j,v_j)$. In contrast to the cosmological literature which considers systems of fixed number density, we consider a system of fixed total mass $M = Nm$. Rearranging gives
\begin{equation}
    H = \frac{M}{N} \left[\frac{1}{2}\sum_{i=1}^N v_i^2 - \frac{1}{2N}\sum_{i\neq j = 1}^N\frac{GM}{|q_i-q_j|}\right],
\end{equation}
which is of the claimed form, up to an irrelevant prefactor.
\subsection{First-order perturbation theory}
A natural way to approximate \eqref{e:density_correlator} is to perform a formal expansion of the generating functional $Z[\bm{J},\bm{K}]$ in powers of the interparticle potential. This perturbative framework is well established in the CSF literature; the required expressions for compact systems involve only minor adjustments and are derived in full in the appendix. To first order in the potential, the $n$-particle density correlator is then given approximately by
\begin{equation}\label{e:fopt}
    \mathbb{E}[\tilde\rho_1(k_1,t_1)\cdots \tilde\rho_n(k_n,t_n)]
    \approx Z[\bm{L}_0,0] - \sum_{i=1}^{n} \int_0^{t_i}\diff t'\int_{k'} v(k',t')
    Z[\bm{L}_0+\bm{L}_i,0] \begin{bmatrix}
    k_i \\ 0
    \end{bmatrix} 
    \cdot
    G(t_i,t') 
    \begin{bmatrix}
    0 \\
    k'
    \end{bmatrix},
\end{equation}
where 
\begin{align}
    \bm{L}_0(u) & := 
    -
    \sum_{i=1}^n 
    \delta_{\rm D}(u-t_i)
    \begin{bmatrix}
    k_i \\ 0
    \end{bmatrix}
    \otimes
    e_i, \label{e:L0} \\
    \bm{L}_i(u) & := 
    -
    \delta_{\rm D}(u-t')
    \begin{bmatrix}
    k' \\ 0
    \end{bmatrix}
    \otimes
    e_{n+1}
    -
    \delta_{\rm D}(u-t')
    \begin{bmatrix}
    -k' \\ 0
    \end{bmatrix}
    \otimes
    e_i. \label{e:Li}
\end{align}
\subsection{Iterated mean-field approximation}
In order to move beyond the perturbative regime, we now employ heuristic reasoning inspired by cosmological structure formation to motivate a non-perturbative approximation scheme. 
The starting point for the mean-field approximation is the characteristic functional $Z[\bm{J}]$ of the single source field. It is straightforward to show that
\begin{equation}
    \mathbb{E}[\tilde\rho_1(k_1,t_1)\cdots \tilde\rho_n(k_n,t_n)]
    = \mathbb{E}\left[e^{\iu \int_0^T\diff u \, \bm{L}_0(u) \cdot \bm{x}(u)}\right].
\end{equation}
Evaluating the integral in the exponent and recalling that $F_{q_i}=0$,
\begin{align}
    \int_0^T\diff u \, \bm{L}(u) \cdot \bm{x}(u)
    & = 
    - \sum_{i=1}^n
    \int_0^T\diff u \, \delta_{\rm D}(u-t_i) 
    \begin{bmatrix}
    k_i \\ 0
    \end{bmatrix}
    \cdot
    \left[G(u,0)x_{0,i} + \int_0^u \diff t' G(u,t') F_i(\bm{x}(t'),t')\right], \\
    & =
    -\sum_{i=1}^n 
    \begin{bmatrix}
    k_i \\ 0
    \end{bmatrix} \cdot  \left[G(t_i,0)x_{0,i} + \int_0^{t_i} \diff t' G(t_i,t') F_i(\bm{x}(t'),t')\right], \\
    & =
    -\sum_{i=1}^n k_i \cdot \left[
    \bar{q}_i(t) + \int_0^{t_i} \diff t'  G_{qp}(t_i,t') F_{p_i}(\bm{x}(t'),t')
    \right],
\end{align}
where $\bar{q}_i(t)$ denotes the inertial trajectory of the $i$th particle in configuration space and where
\begin{equation}\label{e:Fnet}
    F_{p_i}(\bm{x},t) = -\frac{1}{N}\sum_{j=1}^N \nabla v(q_i - q_j,t).
\end{equation}
For simplicity, we now assume that the matrix $G_{qp}(t,t') \in \mathbb{R}^{d\times d}$ is a multiple of the identity matrix, $G_{qp}(t,t') = g_{qp}(t,t') I_d$. In the case of the equal-time density-density correlator $P(k,t)$ we have,
\begin{align}
    \int_0^T\diff u \, \bm{L}_0(u) \cdot \bm{x}(u)
    & = - k \cdot ( \bar{q}_1(t) - \bar{q}_2 (t) ) - k \cdot \int_0^t \diff u \, g_{qp}(t,u) \big[F_{p_1}(\bm{x}(u),u) - F_{p_2} (\bm{x}(u),u))\big].
\end{align}
In the large-$N$ limit we then make the (admittedly heuristic) assumption that the net force on particle 1, given by the average \eqref{e:Fnet}  over the remaining $N-2$ particles, is approximated by the force generated by particle 2. By symmetric reasoning one then obtains,
\begin{align}
    F_{p_1}(\bm{x}(t),t) & \approx - \nabla v \big(q_1(t) - q_2(t) ,t\big), \\
    F_{p_2}(\bm{x}(t),t) & \approx - \nabla v \big(q_2(t) - q_1(t) ,t\big).
\end{align}
Then by parity invariance assumption \eqref{e:parity} we obtain
\begin{equation}
    \nabla v(-q,t) = - \nabla v(q,t).
\end{equation}
Thus we obtain the following approximation for the integral,
\begin{equation}
    \int_0^T\diff u \, \bm{L}_0(u) \cdot \bm{x}(u) \approx - k \cdot \left( \bar{q}_1(t) - \bar{q}_2 (t) \right) + 2 \int_0^t \diff u \, g_{qp}(t,u) \, k\cdot \nabla v \big(q_1(u) - q_2(u) ,u\big).
\end{equation}
Let us denote by $\bar{P}(k,t)$ the equal-time density-density correlator under this approximation,
\begin{align}
    \bar{P}(k,t)
    := \mathbb{E}\left[e^{- \iu k \cdot ( \bar{q}_1(t) - \bar{q}_2 (t) )} e^{2\iu \int_0^t \diff u \, g_{qp}(t,u) \, k\cdot \nabla v (q_1(u) - q_2(u) ,u)}\right].
\end{align}
Then we expect
\begin{equation}
    P(k,t) \underset{N\to\infty}{\approx} \bar{P}(k,t).
\end{equation}
The strategy behind the mean-field approximation is to further approximate $\bar{P}(k,t)$ by replacing the random variable 
\begin{equation}\label{e:RV}
    k\cdot \nabla v (q_1(t) - q_2(t) ,t)
\end{equation}
by a non-fluctuating c-number function $c(k,t)$, thereby defining the mean-field density-density correlator
\begin{equation}
    P_\textrm{mf}(k,t) := P_0(k,t) \, e^{2\iu \int_0^t \diff u \, g_{qp}(t,u) \, c(k,u)}.
\end{equation}
For an appropriately chosen c-number function, we then expect to obtain an  (uncontrolled) approximation of $P(k,t)$ in the sense that
\begin{equation}
    P(k,t) \underset{N\to\infty}{\approx} \bar{P}(k,t) \approx P_\textrm{mf}(k,t).
\end{equation}
A plausible choice of c-number function is the expected value of the random variable \eqref{e:RV}; that is,
\begin{equation}
    c_0(k,t):=\mathbb{E}\big[k \cdot \nabla v \big(q_1(t) - q_2(t) ,t\big)\big].
\end{equation}
Unfortunately, this choice does not produce a useful approximation. The issue is that
\begin{align}
    c_0(k,t)
    & =
    \iu k \cdot \int_{k'} k' v(k') \mathbb{E}[e^{\iu k' q_1(t)} e^{-\iu k' q_2(t)}], \\
    & = \iu k \cdot \int_{k'} k' v(k') P(-k',t), \\
    & = 0,
\end{align}
where we have used the fact that $v$ (and thus $P$) is parity invariant \eqref{e:parity}. Thus, we obtain the uninteresting approximation,
\begin{equation}
    P(k,t) \approx P_\textrm{mf}(k,t)=P_0(k,t).
\end{equation}
In order to motivate a better choice, observe that $c_0$ can be expressed as a certain convolution evaluated at vanishing wave vector,
\begin{equation}
    c_0(k,t) = \frac{k \cdot \big(\mathcal{F}[\nabla v] \ast P \big) (0)}{(2\pi)^d}.
\end{equation}
The above observation suggests considering the convolution evaluated at an arbitrary wave vector $k' \in \mathbb{R}^d$,
\begin{equation}
    c_{k'}(k,t) = \frac{k \cdot \big(\mathcal{F}[\nabla v] \ast P \big) (k')}{(2\pi)^d}.
\end{equation}
Following the literature on CSF \cite{bartelmann2021kinetic}, we propose to choose $k'=k$, which corresponds to the following c-number replacement of the random variable \eqref{e:RV}
\begin{equation}
    c_k(k,t)
    = \iu k \cdot \int_{k'} k' v(k',t) P(k - k',t).
\end{equation}
The resulting mean-field density-density correlator satisfies the following nonlinear integral equation,
\begin{align}
    P_{\rm mf}(k,t)
    & = 
    P_0(k,t)\exp\left[- 2 k \cdot \int_0^t \diff t' g_{qp}(t,t') \int_{k'} k' v(k',t') 
    P_{\rm mf}(k - k',t')\right],
\end{align}
which we abbreviate as the functional equation
\begin{equation}\label{e:mft}
    P_{\rm mf} = T[P_{\rm mf}].
\end{equation}
Although solving the functional equation \eqref{e:mft} is a non-trivial task, one can obtain an approximate solution by a heuristic iteration method, which we call iterated mean-field theory (MFT$_n$). Specifically, starting with the initial guess $P_0$ we form a sequence of functions $\{ P_n \}_{n\geq 0}$ defined by the recursion $P_{n+1} = T[P_n]$.
It is crucial to emphasize the distinction between the sequence of iterates $\{P_n \}_{n\geq 1}$ and the mean-field correlator $P_\textrm{mf}$. In particular, the sequence $\{P_n \}_{n\geq 1}$ has no guarantee of convergence. It is worth remarking, however, that if the map $T : P \longmapsto T[P]$ is a contraction mapping, then the convergence result
\begin{equation}
    \lim_{n\to\infty} P_n = P_\textrm{mf}
\end{equation}
follows from the Banach fixed-point theorem. It would be interesting to explore under what conditions, if any, the contractive property is satisfied. Finally, we comment that error of MFT$_n$ can be quantified in terms of the difference of iterates, using  the fact that the residual function $r_n$ satisfies the identity
\begin{align}\label{e:residual}
    r_n & := P_n - T[P_n], \\
    & = P_n - P_{n+1}.
\end{align}

\section{Examples}\label{sec:examples}
In the remainder of the paper we explore the first-order perturbation theory (FOPT) and the iterated mean-field approximation (MFT$_n$) in a number of analytically tractable examples. In order to facilitate comparison, we focus on the equal-time density-density correlator \eqref{e:P(k,t)}. The notation $P_n$ is used to denote the $n$th iteration of MFT$_n$ and $P_\textrm{pert}$ to denote the FOPT result.
\subsection{Short-ranged interactions}
Consider a system of non-relativistic particles\footnote{A related problem has been investigated in \cite{kozlikin2023ultracold}.} of mass $m=1$ moving in $M=\mathbb{R}$ with initial positions and momenta drawn from the Gaussian distributions  $q_i \sim N(0,\sigma_q^2)$ and $p_i \sim N(0,\sigma_p^2)$. Clearly, the initial conditions break the homogeneity in configuration space. The single-particle Green function for this simple problem is the following $2\times 2$ matrix
\begin{equation}\label{e:green_NR}
    G(t,t')=
    \begin{bmatrix}
    1 & t - t' \\
    0 & 1
    \end{bmatrix}.
\end{equation}
The interparticle potential is also chosen to be Gaussian, normalized such that it approaches a delta function $v(q,t) \to g \delta_{\rm D}(q)$ in the limit $\sigma \to 0$,
\begin{equation}
    v(q,t) = \frac{g}{\sqrt{2\pi\sigma^2}} e^{-\frac{1}{2\sigma^2}q^2} \implies
    v(k,t) = g e^{-\frac{1}{2}\sigma^2k^2}.
\end{equation}
The inertial correlator is thus given by
\begin{align}
    P_0(k,t)
    = \mathbb{E} \left[e^{-\iu k(q_1 + t p_1) + \iu k(q_2 + t p_2)}\right]
    = e^{-k^2(\sigma_q^2 + t^2 \sigma_p^2)}.
\end{align}
The elementary nature of this problem makes it an ideal testing ground for comparison of FOPT and MFT$_n$. Let us first consider first-order perturbation theory. Recalling \eqref{e:fopt}, we obtain
\begin{align}
    P_\textrm{pert}(k,t)& = Z[\bm{L}_0,0] +
    k \int_{0}^t\diff t'(t-t')\int_{k'} k' ge^{-\frac{1}{2}\sigma^2k'^2}
    \big(
    Z[\bm{L}_0 + \bm{L}_2,0]
    -
    Z[\bm{L}_0 + \bm{L}_1,0]
    \big),
\end{align}
where
\begin{align}
    Z[\bm{L}+\bm{L}_1,0]
    & = \mathbb{E} \left[e^{-\iu k(q_1 + t p_1) + \iu k(q_2 + t p_2) + \iu k'(q_1+t'p_1) - \iu k'(q_3 + t' p_3)}\right], \\
    & = e^{-(k^2+k'^2-kk') \sigma_q^2 -(k^2t^2+k'^2t'^2-kk'tt') \sigma_p^2}, \\
    Z[\bm{L}+\bm{L}_2,0]
    & = \mathbb{E} \left[e^{-\iu k(q_1 + t p_1) + \iu k(q_2 + t p_2) + \iu k'(q_2+t'p_2) - \iu k'(q_3 + t' p_3)}\right], \\
    & = e^{-(k^2+k'^2+kk') \sigma_q^2 -(k^2t^2+k'^2t'^2+kk'tt') \sigma_p^2}.
\end{align}
Carrying out the Fourier integral we obtain,
\begin{align}
    P_\textrm{pert}(k,t)
    & = P_0(k,t)
    \left\{
    1 -
    \frac{g k^2}{\sqrt{\pi/2}} 
    \int_0^t\diff t'(t-t')
    \frac{\alpha(t,t')}{[\sigma^2 + 2\alpha(t',t')]^{3/2}}
    \exp\left[
    \frac{1}{2}k^2
    \frac{\alpha(t,t')^2}{\sigma^2 + 2\alpha(t',t')}
    \right]
    \right\},
\end{align}
where we have defined
\begin{equation}
    \alpha(t,t') := \sigma_q^2 + \sigma_p^2 tt'.
\end{equation}
The remaining integral over $t'$ requires numerical evaluation. Notice that the first-order correction to the inertial correlator $P_0(k,t)$ is strictly negative, which jeopardizes the positivity constraint $P(q,t) > 0$, satisfied by the configuration-space correlator \eqref{e:P(q,t)}.

\begin{figure}
\small
\centering
\includegraphics[width=0.4\textwidth]{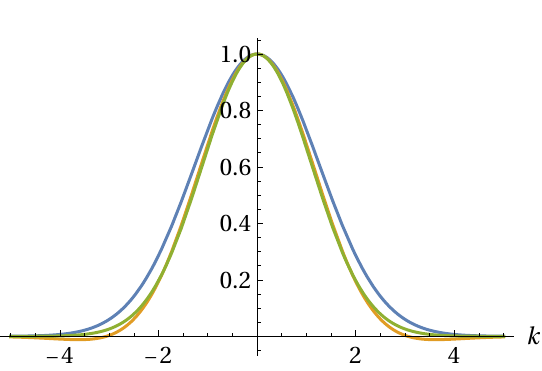}
\includegraphics[width=0.4\textwidth]{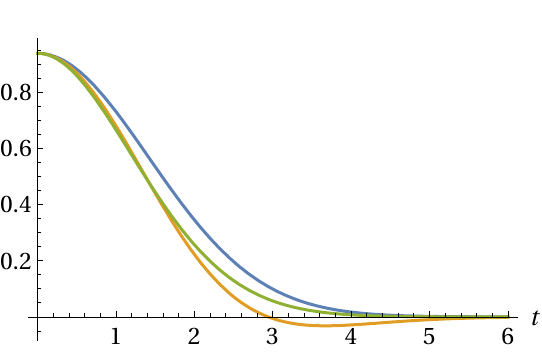}
\caption{Cross-sections of the density-density correlator for short-ranged potential with range $\sigma>0$ along $t=1$ (left) and $k=1$ (right). Shown is the inertial approximation $P_0$ (blue), the first-order perturbation improvement $P_\textrm{pert}$ (orange) and the first iteration of the mean-field approximation $P_1$ (green). The interparticle potential and initial phase-space density parameters are $g=0.1$, $\sigma=0.01$, $\sigma_q=0.25$, $\sigma_p=0.5$.\label{fig:FOPT_vs_MFT}}
\end{figure}

Next we consider the $\textrm{MFT}_1$ approximation,
\begin{align}
    P_1(k,t)
    & = P_0(k,t)\exp\left[-2gk \int_0^t \diff t'(t-t')\frac{1}{2\pi}\int\diff k' k' e^{-\frac{1}{2}\sigma^2 k'^2} P_0(k-k',t')\right].
\end{align}
Similar to FOPT, the Fourier integral can be carried out analytically, while the $t'$ integral requires numerical evaluation
\begin{equation}
    \frac{1}{2\pi}\int\diff k' k' e^{-\frac{1}{2}\sigma^2 k'^2} P_0(k-k',t') \\
    = \frac{k}{\sqrt{\pi/2}}\frac{\alpha(t',t')}{[\sigma^2+2\alpha(t',t')]^{3/2}} \exp\left[-\sigma^2k^2\frac{\alpha(t',t')}{\sigma^2+2\alpha(t',t')}\right].
\end{equation}
Now we compare FOPT and $\textrm{MFT}_1$. Since FOPT is perturbative in $g$, we should consider the linear term in the expansion of $P_1(k,t)$ about $g=0$. 
\begin{align}
    P_1(k,t)
    & = P_0(k,t)
    \left\{
    1 -
    \frac{2gk^2}{\sqrt{\pi/2}}\int_0^t\diff t' (t-t') \frac{\alpha(t',t')}{[\sigma^2+2\alpha(t',t')]^{3/2}} \exp\left[-\sigma^2k^2\frac{\alpha(t',t')}{\sigma^2+2\alpha(t',t')}\right]
    \right\} + O(g^2).
\end{align}
Now consider the relative difference $\delta(k,t)$ between the $O(g)$ terms in $P_\textrm{pert}(k,t)$ and $P_1(k,t)$,
\begin{align}
    \delta(k,t)
    & = 2\frac{\int_0^t\diff t' (t-t') \frac{\alpha(t',t')}{[\sigma^2+2\alpha(t',t')]^{3/2}} \exp\left[-\sigma^2k^2\frac{\alpha(t',t')}{\sigma^2+2\alpha(t',t')}\right]}{\int_0^t\diff t'(t-t')
    \frac{\alpha(t,t')}{[\sigma^2 + 2\alpha(t',t')]^{3/2}}
    \exp\left[
    \frac{1}{2}k^2
    \frac{\alpha(t,t')^2}{\sigma^2 + 2\alpha(t',t')}
    \right]} - 1.
\end{align}
Recall that FOPT is expected to be accurate at early times, where the interacting trajectories are closely approximated by their inertial values. In this case, the time integrals defining $\delta$ can be approximated at leading order in $t$ giving,
\begin{align}
    \delta(k,t)
    & \approx 2 \exp\left[k^2\left(\sigma_p^2t^2+\frac{3\sigma_q^4}{2(\sigma^2+2\sigma_q^2)}\right)\right] - 1, \\
    & \geq 1.
\end{align}
The above bound establishes, in a concrete model, that $\textrm{MFT}_1$ introduces uncontrollable errors at early times, which is precisely the regime where FOPT is expected to be applicable\footnote{It does not, however, speak to $\textrm{MFT}_{n\geq 2}$, nor the non-perturbative solution of the integral equation \eqref{e:mft}.}. It is not clear if these errors are practically relevant, however, since the absolute error is vanishing as $t \to 0$. On the other hand, we expect FOPT to break down at late times when the inertial and interacting trajectories diverge. The equal-time density-density correlator in the FOPT and $\textrm{MFT}_1$ approximation is illustrated in Fig.~\ref{fig:FOPT_vs_MFT}. Despite the error of $\textrm{MFT}_1$ at early times, it is reasonable to expect $\textrm{MFT}_{n\geq 1}$ to provide a much better description of the physics at late times than FOPT.

\subsubsection{Contact interaction limit}
In order to make further analytical progress, we consider the MFT$_n$ in the limit $\sigma\to 0$ (delta-function potential). In the case of $\textrm{MFT}_1$, the $t'$ integral can now be carried out producing the following Gaussian correlator,
\begin{align}
    P_1(k,t) = e^{-\frac{1}{2}\Sigma(t)k^2} \implies
    P_1(q,t) = \int_k e^{\iu k q} P_1(k,t) = \frac{1}{\sqrt{2\pi \Sigma(t)}}e^{-\frac{1}{2\Sigma(t)}q^2},
\end{align}
where
\begin{equation}
    \Sigma(t) := 2 \alpha(t,t) +
    \frac{2g}{\sqrt{\pi}\sigma_p^2}
    \left[\sigma_q-\sqrt{\alpha(t,t)}+\frac{\sigma_pt}{2}\log\left(1+\frac{2\sigma_pt\big(\sqrt{\alpha(t,t)}+\sigma_pt\big)}{\sigma_q^2}\right)\right].
\end{equation}
Inspecting $P_1(q,t)$ above, we notice that unlike for FOPT, the $\textrm{MFT}_1$ approximation satisfies strict positivity, adding to the plausibility that $\textrm{MFT}_1$ is applicable at late times.
Having determined $P_1(k,t)$ in closed form, we attempt to understand the implications for the physics at late times. If the interactions are repulsive ($g>0$) then $\Sigma$ grows monotonically with time and the correlator $P_1(q,t)$ spreads. If the interactions are attractive ($g < 0$), then the term proportional to $g$ competes and $P_1(q,t)$ can either spread out (weak interaction), undergo a bounce (moderate interaction) or collapse to a delta function (strong interaction). If blowup occurs then the time of blowup can be estimated by Taylor expanding $\Sigma$ assuming small velocity dispersion ($\sigma_p \ll 1$),
\begin{equation}
    \Sigma(t) = \sigma_q^2 + \frac{gt^2}{2\sqrt{\pi} \sigma_q} + O(\sigma_p^2).
\end{equation}
Thus, $P_1(q=0,t)$ blows up at a time determined by the solution of $\Sigma(t_{\rm c}) = 0$,
\begin{equation}
    t_{\rm c} \approx 1.88\sqrt{\frac{\sigma_q^3}{-g}}.
\end{equation}

\begin{figure}
\small
\centering
\includegraphics[width=0.4\textwidth]{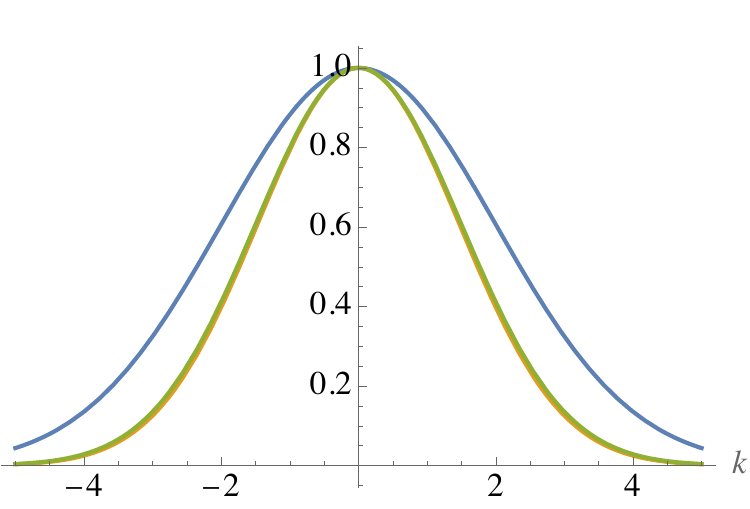}
\includegraphics[width=0.4\textwidth]{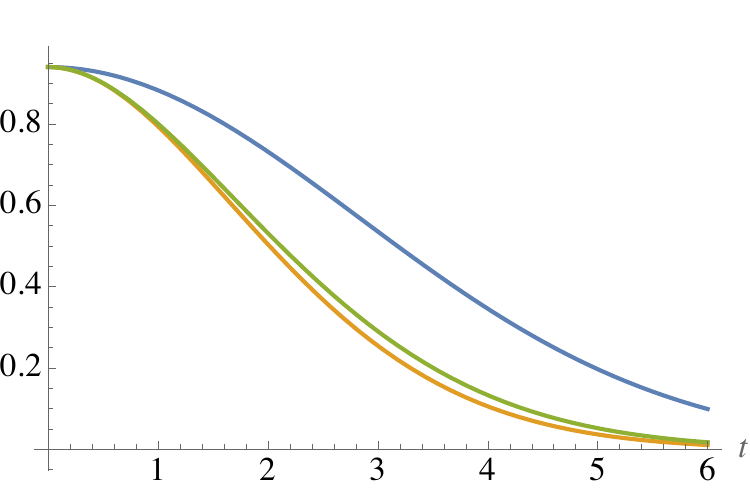}
\caption{Cross-sections of the iterated mean-field density-density correlator $P_n(k,t)$ along $t=1$ (left) and $k=1$ (right) in the delta function potential limit $\sigma \to 0$. Shown are $n=0$ (blue), $n=1$ (orange) and $n=2$ (green). The interparticle potential and initial phase-space density parameters are $g=0.1$, $\sigma_q=0.25$, $\sigma_p=0.5$.\label{fig:MFTn}}
\end{figure}

\begin{figure}
\small
\centering
\includegraphics[width=0.4\textwidth]{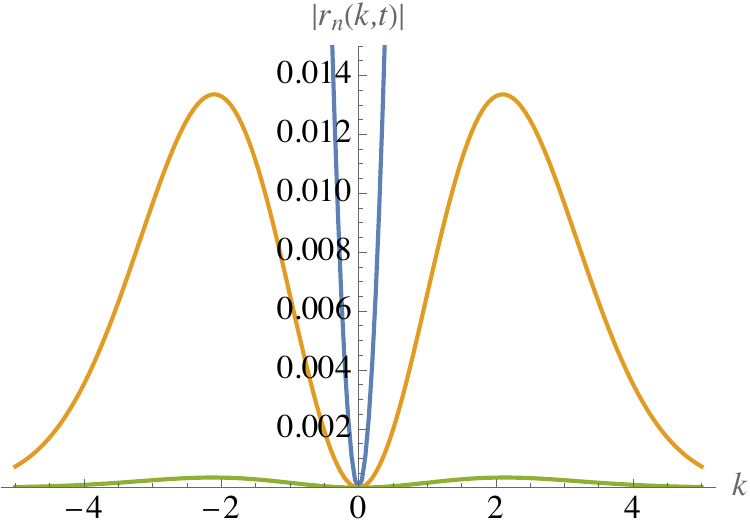}
\includegraphics[width=0.4\textwidth]{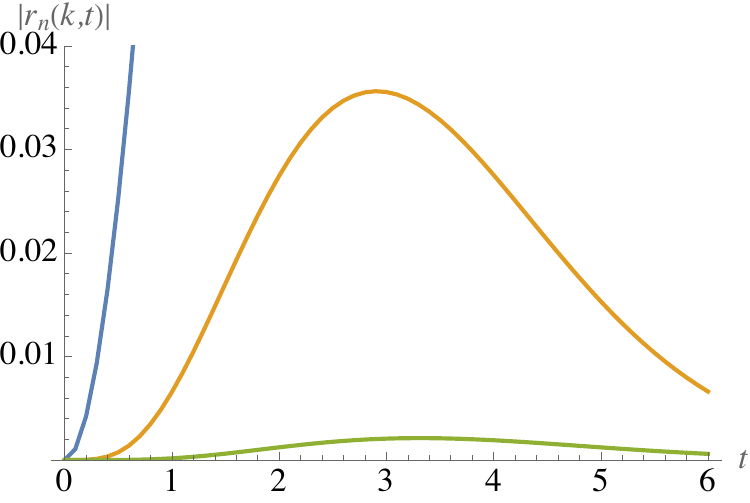}
\caption{Cross-sections of the absolute pointwise residual \eqref{e:residual} along $t=1$ (left) and $k=1$ (right) in the delta function potential limit $\sigma \to 0$. Shown are $n=0$ (blue), $n=1$ (orange) and $n=2$ (green). The interparticle potential and initial phase-space density parameters are $g=0.1$, $\sigma_q=0.25$, $\sigma_p=0.5$.\label{fig:residual}}
\end{figure}

It is also possible to  explore $\textrm{MFT}_2$ using a semi-analytical approach. In particular, the $k'$ integral defining $P_2(k,t)$ can be carried out in closed form, leaving a $t'$ integral, which requires numerical evaluation (see Fig.~\ref{fig:MFTn}). For $n\geq 3$, however, numerical integration over the region $(k',t') \in \mathbb{R}\times [0,t]$ is required. In Fig.~\ref{fig:residual}, we plot the pointwise residual \eqref{e:residual}, providing evidence that $\textrm{MFT}_2$ significantly improves the approximation of the mean-field correlator compared to $\textrm{MFT}_1$. The price paid for the improved accuracy is a loss of analytical control.

\subsubsection{Extension to three dimensions}
Now we briefly discuss the generalization to $M=\mathbb{R}^3$. Assuming that $v$ and $P_0$ are spherically symmetric functions, we overload notation by expressing their dependence on the norm $k:=|\vec k|$ as $v(\vec k,t) = v(k,t)$ and $P_0(\vec k ,t)=P_0(k,t)$. In the $\textrm{MFT}_1$ approximation,
\begin{align}
    P_1(\vec k,t)
    & = P_0(\vec k,t)\exp\left[-2 \vec k \cdot \int_0^t \diff t'(t-t')\int_{k'} \vec k' v(\vec k',t') P_0(\vec k-\vec k',t')\right], \\
    & = P_0(k,t)\exp\left[-2k \int_0^t \diff t'(t-t')\frac{1}{(2\pi)^2}\int_{0}^\infty \diff k'  k'^3 v(k',t') \int_{-1}^1\diff x \, x \, P_0\left({\sqrt{k^2 + k'^2 - 2kk'x}},t'\right)\right].
\end{align}
Let $v(k,t) = g e^{-\frac{1}{2}\sigma^2 k^2}$, $q_i\sim N(0,\sigma_q^2I_3)$, $p_i\sim N(0,\sigma_p^2I_3)$. Performing the $k'$ integral, sending $\sigma \to 0$, then performing the $x$ integral followed by the $t'$ integral gives 
\begin{align}
    P_1(q,t)
    & = 
    \frac{1}{\sqrt{(2\pi \Sigma(t))^3}} e^{-\frac{1}{2\Sigma(t)} q^2}, \\
    \Sigma(t)
    & = 2 \left[\alpha(t,t) + \frac{g}{4\pi^{3/2}}\frac{\sqrt{\alpha(t,t)} - \sigma_q}{\sigma_q^2\sigma_p^2}\right],
\end{align}
which exhibits the same qualitative behavior as $d=1$.

\subsection{Gravitating sheet model}
In the next example we attempt to use $\textrm{MFT}_1$ to understand the late time physics of the gravitating sheet model (GSM), which can be regarded as a system of non-relativistic particles of mass $m=1$ in $M=\mathbb{R}$ with the following interparticle potential,
\begin{equation}
    v(q,t) = \frac{g}{2} |q| e^{-\alpha |q|} \implies
    v(k,t) = -g \frac{k^2-\alpha^2}{(k^2+\alpha^2)^2},
\end{equation}
where $g>0$ is proportional to the areal mass density of the sheet and  $\alpha > 0$ is a regularization parameter required to ensure convergence of the Fourier integrals. For the purposes of analytical evaluation, we take the sheets to be initially Laplace distributed in phase space; that is, $q_i \sim \Laplace(0,b_q)$ and $p_i \sim \Laplace(0,b_p)$. Then we obtain\footnote{Recall that the probability density function for the Laplace distribution with mean $\mu \in \mathbb{R}$ and diversity $b>0$ is given by
$
    f(x \, | \, \mu, b) = \frac{1}{2b}e^{-\frac{|x-\mu|}{b}}
$.}
\begin{align}
    P_0(k,t)
    & := \mathbb{E} \left[e^{-\iu k(q_1 + t p_1) + \iu k(q_2 + t p_2)}\right], \\
    & = \frac{1}{(1+b_q^2k^2)^2(1+b_p^2t^2k^2)^2}.
\end{align}
Consider $\textrm{MFT}_1$,
\begin{align}
    P_1(k,t)
    & = P_0(k,t)\exp\left[-2k \int_0^t \diff t'(t-t')\frac{1}{2\pi}\int\diff k' k' v(k') P_0(k-k',t')\right].
\end{align}
Convergence requires careful attention to the order of operations. First performing the $k'$ integral, then letting $\alpha \to 0$ and finally carrying out the $t'$ integral we obtain
\begin{align}
    P_1(k,t) =
    \frac{1}{(1+b_q^2k^2)^2(1+b_p^2t^2k^2)^2}
    \exp
    \left[
    \frac{g k^2 t \left(\frac{\log \left(1+b_p^2t^2 k^2\right)}{k^2}-\frac{b_p t b_q^2 \left(1+b_q^2 k^2\right)}{b_p t+b_q}+2 b_q^2 \left(2+b_q^2 k^2\right) \log \left(1+\frac{b_p t}{b_q}\right)\right)}{4 b_p \left(1+b_q^2 k^2\right)^2}
    \right].
\end{align}
A first observation is that the time dependence of
$P_1(k,t)$ undergoes a phase transition from algebraic decay at early times to exponential growth at late times. An estimate of the transition time $t_{\rm c}$ can be made by considering a broad spatial distribution function ($b_q \gg 1$). Expanding the argument of the exponential in powers of $\frac{1}{b_q}$ one finds the following time dependence at leading order,
\begin{align}
    P_1(k,t) \propto \frac{e^{t^2/(4 b_q)}}{(1+b_p^2t^2k^2)^2}.
\end{align}
Then solving for the stationary point
\begin{equation}
    \frac{\partial P_1}{\partial t}(k,t_{\rm c}) = 0,
\end{equation}
we obtain
\begin{align}
    t_{\rm c} = \frac{1}{b_pk}\sqrt{\frac{8 b_q b_p^2 k^2}{g} - 1}
    \approx
    \sqrt{\frac{8b_q}{g}}.
\end{align}
At this point we recall that the dispersion of $\Laplace(0,b_q)$ is given by $\sigma = \sqrt{2}b_q$ and the free-fall time in the potential $\frac{g}{2}|q|$ is given by $\sqrt{4q/g}$. Thus, the transition time $t_{\rm c}$ coincides with the free-fall time starting at $\sqrt{2} \approx 1.4$ deviations from the mean.

Next we argue that in fact the transition time corresponds to the breakdown of the mean-field approximation in this model. In particular, once $t$ reaches a critical value, the configuration-space correlator $q \mapsto P_1(q,t)$ develops negative lobes. For simplicity, consider the limit of vanishing velocity dispersion ($b_p \to 0$), so that
\begin{align}
    P_1(k,t) 
    & = 
    \frac{1}{(1 + b_q^2 k^2)^2}
    \exp
    \left[
    \frac{g b_qk^2(3+b_q^2k^2)}{(1+4b_q^2k^2)^2}
    t^2
    \right].
\end{align}
Expanding around $k=0$,
\begin{equation}
    P_1(k,t) = 1 + b_q \left(\frac{3}{4} gt^2 - 2 b_q\right)k^2 + O(k^4),
\end{equation}
which shows that the maximum at $k=0$ bifurcates into two maxima for $t \geq \sqrt{8b_q/(3g)}$ located at $k=\pm k_\ast$. Thus, assuming $b_q \gg 1$, the time of bifurcation is within $O(1)$ factors of $t_{\rm c}$. Now consider the configuration-space correlator
\begin{equation}
    P_1(q,t) = \frac{1}{2\pi}\int_{\mathbb{R}}\diff k \, e^{\iu k q} P_1(k,t).
\end{equation}
For $t \geq t_{\rm c} \gg 1$, the above integral can be estimated by the Laplace method. The saddle points at $k=\pm k_\ast$ contribute terms of the form $e^{\pm \iu k_\ast q}$, which combine to yield
\begin{align}
    P_1(q,t)
    \propto \cos(q k_\ast),
\end{align}
which changes sign at $q = \pm \frac{\pi}{2k_\ast}$.
\subsection{Criticality in the gravitating sheet model}
\begin{figure}
\small
\centering
\includegraphics[width=0.75\textwidth]{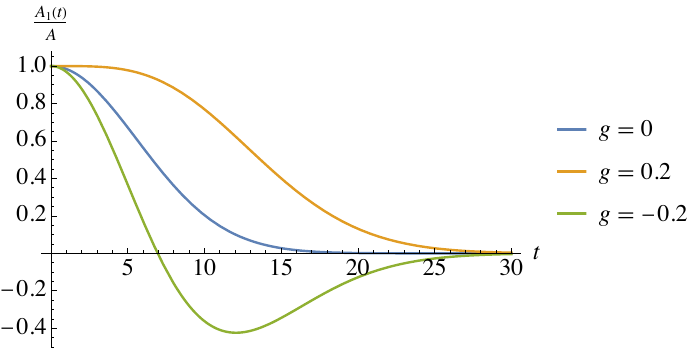}
\caption{Time development of $A_1(t)$ for different choices of interaction coupling $g$. The exact solution for the system of free particles $A_1(t)/A=e^{-\frac{1}{2}k_1^2\sigma^2 t^2}$ is shown in blue. The development for the gravitational (orange) and electrostatic (green) interacting systems was computed using first-order perturbation theory. The system size is $L=2\pi$ and the velocity dispersion has been tuned to the critical value for the gravitational instability $\sigma^2=\sigma_{\rm cr}^2\approx0.032$. \label{fig:A1}}
\end{figure}

As a final application, we use FOPT to analyze criticality in the GSM on the interval $[0,L]$ with periodic boundary conditions. The potential is now
\begin{align}
    v(q) & = \frac{g}{2}\left(|q|-\frac{q^2}{L}\right).
\end{align}
Converting to Fourier space,
\begin{align}
    v(k_n) & := \int_0^L \diff q \, e^{-\iu k_n q} v(q), \\
    & =
    \frac{g}{2}
    \begin{cases}
    -\frac{L^2}{2\pi^2n^2}, & n\neq 0 \\
    \frac{L^2}{6} & n = 0
    \end{cases},
\end{align}
where $k_n = \frac{2\pi}{L}n$ and $n \in \mathbb{Z}$. 
Consider the initial phase-space density $\rho_0 = \bigotimes_{j=1}^N f_0$, where
\begin{align}
    f_0(q,p) & =
    \frac{1 + A \cos (k_1 q)}{L} 
    \frac{e^{-\frac{p^2}{2\sigma^2}}}{\sqrt{2\pi\sigma^2}},
\end{align}
and $|A|<1$.
The first-order improvement to the single-particle density correlator is
\begin{align}
    \mathbb{E}[\rho_1(k_n,t)] & = Z[\bm{L}_0,0] -
    k_n \int_{0}^t\diff t'(t-t') \left[\frac{1}{L}\sum_{n' \in \mathbb{Z}} k_{n'} v(k_{n'},t')
    Z[\bm{L}_0 + \bm{L}_1,0]\right] + O(g^2),
\end{align}
where
\begin{align}
    Z[\bm{L}_0,0]
    & = \mathbb{E}[e^{-\iu k_n (q_1+t p_1)}], \\
    & = \left[\delta_{n} + \frac{A}{2}(\delta_{n+1}+\delta_{n-1})\right]e^{-\frac{1}{2}\sigma^2k_n^2t^2}, \\
    Z[\bm{L}_0 + \bm{L}_1,0]
    & = \mathbb{E}[e^{-\iu k_n (q_1+t p_1)+\iu k_{n'}(q_1 + t'p_1)-\iu k_{n'}(q_2+t'p_2)}], \\
    & = \left[\delta_{n'-n} + \frac{A}{2}(\delta_{n'-n+1}+\delta_{n'-n-1})\right]\left[\delta_{n'} + \frac{A}{2}(\delta_{n'+1}+\delta_{n'-1})\right]
    e^{-\frac{1}{2}\sigma^2(k_nt-k_{n'}t')^2}
    e^{-\frac{1}{2}\sigma^2k_{n'}^2t'^2}.
\end{align}
The result is
\begin{align}
    \mathbb{E}[\rho_1(q,t)]
    & = \frac{1}{L}\sum_{n\in\mathbb{Z}}e^{\iu k_n q}\,\mathbb{E}[\rho_1(k_n,t)], \\
    & = \frac{1+A\cos(k_1 q)e^{-\frac{1}{2}k_1^2\sigma^2 t^2}}{L} + O(g), \\
    & = \frac{1+A_1(t)\cos(k_1 q)+A_2(t)\cos(k_2q)}{L} + O(g^2),
\end{align}
where $A_1(t)$ and $A_2(t)$ can be computed in closed form. The first few terms of their Taylor expansions are given by
\begin{align}
    \frac{A_1(t)}{A} & = 1+\frac{gL -4\pi^2\sigma^2}{2L^2}t^2 + \frac{2 \pi ^2 \sigma ^2 \left(3 \pi ^2 \sigma ^2-g L\right)}{3 L^4}t^4 + O(t^6), \label{e:A1}\\
    \frac{A_2(t)}{A^2} & = \frac{g}{2L}t^2 + O(t^4).
\end{align}
The expansion for $A_1$ reveals distinct qualitative behaviors for the electrostatic ($g<0$) and gravitational ($g>0$) system. In the electrostatic case, interactions evidently expedite the decay of $A_1(t)$ (see Fig.~\ref{fig:A1}), while in the gravitational problem, the quadratic term in the Taylor expansion of $A_1(t)$ indicates an instability for velocity dispersion below a critical value $\sigma^2 \leq \sigma_{\rm cr}^2$ where
\begin{equation}
    \sigma_{\rm cr}^2 := \frac{g}{k_1^2L}.
\end{equation}
The above result agrees with the critical point derived from linear stability analysis of the Vlasov-Poisson system \cite[Eq.~(11)]{ivanov2001critical}.

In the gravitational system with $\sigma^2 < \sigma_{\rm cr}^2$, the amplitude $A_1(t)$ reaches a maximum at some $t_{\rm sat} > 0$. It has been argued that the amplitude at saturation $A_{\rm sat}:=A_1(t_{\rm sat})$ should be considered as the order parameter for a dynamical phase transition \cite{ivanov2001critical, ivanov2005criticality}. The scaling of the order parameter with the control parameter,
\begin{equation}
    \theta := \frac{\sigma^2-\sigma_{\rm cr}^2}{\sigma_{\rm cr}^2},
\end{equation}
has been determined by numerically solving the Vlasov-Poisson system and found to have a universal critical exponent $\beta = 1.995 \pm 0.0034$ \cite{ivanov2005criticality},
\begin{equation}
    \frac{A_{\rm sat} - A}{A} 
    \mathrel{\propto}
    \begin{cases}
    (-\theta)^\beta, &  \theta < 0 \\
    0, & \theta \geq 0
    \end{cases}
    \quad\quad\quad
    \textrm{,}
    \quad\quad\quad
    |\theta| \ll 1. 
\end{equation}
In first-order perturbation theory, one can estimate the time of saturation from the solution of $A_1'(t_{\rm sat}) = 0$ using the fourth-order truncated Taylor expansion  \eqref{e:A1}. One finds a universal critical exponent of $\beta=2$ and a universal prefactor of $1.5$,
\begin{equation}
    \frac{A_{\rm sat} - A}{A}
    =
    \begin{cases}
    \frac{3}{2}\theta^2, & \theta < 0 \\
    0, & \theta \geq 0
    \end{cases}
    \quad\quad\quad
    \textrm{,}
    \quad\quad\quad
    |\theta| \ll 1.
\end{equation}

\section{Discussion}\label{sec:discussion}

In the case of short-ranged interactions, our comparison of FOPT and MFT$_n$ reveals a clear division of validity regimes.  At early times, when particle trajectories remain close to their inertial paths, FOPT provides an accurate description of the density-density correlator.  As interactions accumulate and inertial and true trajectories diverge, however, the perturbative expansion quickly loses its positivity and physical plausibility.  By contrast, the mean-field approximation in its first iteration (MFT$_1$) preserves positivity and captures the broadening or collapse of the correlator at late times, even in the singular contact‐interaction limit $\sigma\to0$, where it becomes analytically tractable.  In that limit, $P_1(q,t)$ remains a well‐behaved Gaussian whose variance $\Sigma(t)$ encodes repulsive spreading for $g>0$ or collapse and possible re‐expansion for $g<0$. The analytical method enables a computation of the blowup time in the attractive regime, which agrees with the expected time of singular collapse.  Numerical evidence further shows that the second mean-field iteration, MFT$_2$, substantially reduces the residual error, indicating that higher iterates may plausibly converge toward the nonperturbative solution of the integral equation $P_{\rm mf}=T[P_{\rm mf}]$. 

Turning to the gravitating sheet model, we find that the mean‐field approximation successfully predicts the onset of collapse in the infinite‐volume limit but that it breaks down once non‐positivity appears in the Fourier‐transformed correlator.  The time scale for this breakdown coincides with the classical free‐fall time up to $O(1)$ factors.  In a finite, periodic domain the first‐order perturbative treatment recovers the linear gravitational instability threshold in perfect agreement with Vlasov-Poisson analysis, reproducing the critical velocity dispersion.  Moreover, the growth of the first Fourier mode exhibits the characteristic saturation behavior of a dynamical phase transition, with a critical exponent $\beta=2$.

In terms of future directions, establishing rigorous convergence criteria for the mean‐field map $T$ remains an important avenue for future work. 
In our exactly solvable examples, MFT$_1$ was shown to break down at short times in the short-range interaction model and at late times in the self‐gravitating sheet model. It will be interesting to determine whether higher-order iterations of the mean‐field scheme can cure these pathologies. In addition, it is important to extend the formalism to physically relevant systems with long‐range interactions in higher dimensions, including realistic astrophysical models or models with velocity‐dependent forces.

\acknowledgments

The authors of this paper are listed alphabetically. The work of M.B. was funded in part by the Deutsche Forschungsgemeinschaft (DFG, German Research Foundation) under Germany’s Excellence    Strategy EXC 2181/1 - 390900948 (the Heidelberg STRUCTURES Excellence Cluster).

\newpage

\appendix

\section{Derivation of Green function}

Consider a Hamiltonian system with state variable $x\in\mathbb{R}^{2d}$, Hamiltonian function $H=H(x,t)$ and Poisson matrix $J \in \mathbb{R}^{2d \times 2d}$ such that $J^2 = -I_{2d}$,
\begin{align}
    H(x,t) & = H_0(x,t) + H_1(x,t), \\
    H_0(x,t) & = \frac{1}{2} x^T h(t) x, \\
    J
    & =
    \begin{bmatrix}
    0 & I_d \\
    -I_d & 0
    \end{bmatrix}.
\end{align}
\begin{remark}
The solution of the initial value problem
\begin{equation}
    \left\{
    \begin{aligned}
    \dot{x}(t) & = J \nabla H(x(t),t) \\
    x(0) & = x
    \end{aligned}
    \right.,
\end{equation}
satisfies
\begin{align}\label{e:recursive}
    x(t) & = G(t,0)x + \int_0^t \diff t' \, G(t,t') J \nabla H_1 (\bar x(t'),t'),
\end{align}
where
\begin{align}\label{e:Green_single}
    G(t,t') & := \mathsf{T}\exp\left[J\int_{t'}^t \diff u \, h(u)\right].
\end{align}
\end{remark}
\begin{proof}
By the Leibniz integral rule,
\begin{align}
    \dot{x}(t)
    & =
    \frac{\partial G(t,0)}{\partial t} x
    +
    \int_0^t \diff t' \frac{\partial G(t,t')}{\partial t} J \nabla H_1(t',\bar x(t'))
    +
    G(t,t) \nabla H_1(t,\bar x(t)).
\end{align}
Using $\frac{\partial G(t,t')}{\partial t} = J h(t) G(t,t')$ and $G(t,t) = I_{2d}$ we obtain
\begin{align}
    \dot{x}
    & =
    Jh(t) \left[G(t,0)x
    +
    \int_0^t \diff t' \, G(t,t') J \nabla H_1(t',\bar x(t'))\right]
    + 
    J \nabla H_1(t, \bar x(t)), \\
    & =
    J h(t) x(t) + J \nabla H_1(t,\bar x(t)), \\
    & = J \nabla H(t,\bar x(t)).
\end{align}
\end{proof}

\begin{example}
Consider the harmonic oscillator
\begin{equation}
    H(x,t) = \frac{\vec p^2}{2m(t)} + \frac{1}{2} k(t) \vec q^2.
\end{equation}
Then the Hamilton equations are $\dot{x}(t) = A(t) x(t)$ where
\begin{align}
    A(t) & =
    \begin{bmatrix}
    0 & \frac{1}{m(t)} \\
    -k(t) & 0
    \end{bmatrix},
\end{align}
and thus the commutator is
\begin{align}
    [A(t),A(t')]
    & =
    \left[\frac{k(t)}{m(t')}- \frac{k(t')}{m(t)}\right]
    \begin{bmatrix}
    I_d & 0 \\
    0 & -I_d
    \end{bmatrix}.
\end{align}
If we set $k=0$, then the commutator vanishes $[A(t),A(t')] = 0$ and thus the time-ordered exponential turns into a regular matrix exponential,
\begin{align}
    G(t,t') 
    & = \mathsf{T}\exp\left[\int_{t'}^t \diff u \, A(u)\right], \\
    & = \exp\left[\int_{t'}^t \diff u \, A(u)\right], \\
    & = I_{2d} + \int_{t'}^t \diff u \, A(u).
\end{align}
\end{example}

\section{Derivation of first-order perturbation theory}
Begin with \eqref{e:n_particle} and use the fact that the interaction Hamiltonian is independent of momentum,
\begin{align}
    & \mathbb{E}[\tilde\rho_1(k_1,t_1)\cdots \tilde\rho_n(k_n,t_n)] \notag \\
    & = \left. \hat \rho_1(k_1,t_1)\cdots \hat \rho_n(k_n,t_n) \exp\left[\int_0^T \diff t'\bm{F}\left(\frac{\delta}{\iu \delta \bm{J}(t')},t'\right) \cdot \frac{\delta}{\delta \bm{K}(t')}\right] Z[\bm{J},\bm{K}]\right|_{\bm{J},\bm{K}=0}, \\
    & = \left. \hat \rho_1(k_1,t_1)\cdots \hat \rho_n(k_n,t_n) \exp\left[\int_0^T \diff t' \sum_{i=1}^N F_{p_i}\left(\frac{\delta}{\iu \delta \bm{J}(t')},t'\right) \cdot \frac{\delta}{\delta K_{p_i}(t')}\right] Z[\bm{J},\bm{K}]\right|_{\bm{J},\bm{K}=0}.
\end{align}
Now,
\begin{align}
    F_{p_i}(\bm{x},t)
    & =
    -\frac{\partial}{\partial q_i} \left[\frac{1}{2N}\sum_{j, k=1}^N v(q_j - q_k,t)\right], \\
    & = -\frac{1}{2N}\sum_{j,k=1}^N \nabla v(q_j - q_k,t) (\delta_{ij} - \delta_{ik}), \\
    & = -\frac{1}{2N}\sum_{j=1}^N \nabla v(q_i - q_j,t) - \nabla v(q_j - q_i), \\
    & = -\frac{1}{N}\sum_{j=1}^N \nabla v(q_i - q_j,t).
\end{align}
Thus,
\begin{align}
    \nabla v(q_i - q_j,t)
    & = \int_{q,q'} \delta_{\rm D}(q-q_i)\nabla v(q-q',t) \delta_{\rm D}(q'-q_j), \\
    & = \int_{q,q'} \left[\int_{k_1} e^{\iu k_1 (q-q_i)}\right]\nabla \left[\int_{k_1} e^{\iu k_2 (q-q')} v(k_2,t)\right]\left[\int_{k_3} e^{\iu k_3 (q'-q_j)}\right], \\
    & = \int_{q,q',k_1,k_2,k_3} (\iu k_2) e^{\iu q (k_1+k_2)} e^{\iu q'(k_3-k_2)} v(k_2,t) e^{-\iu k_1 q_i - \iu k_3 q_j}, \\
    & = (2\pi)^{2d}\int_{k_1,k_2,k_3} (\iu k_2) \delta_{\rm D}(k_1+k_2)\delta_{\rm D}(k_3-k_2) v(k_2,t) e^{-\iu k_1 q_i - \iu k_3 q_j}, \\
    & =
    \iu \int_k e^{\iu k q_i} k \, v(k,t) e^{-\iu k q_j}.
\end{align}
Thus,
\begin{equation}
    F_{p_i}(\bm{x},t)
    =
    -\iu \int_k e^{\iu k q_i} k \, v(k,t) \left[\frac{1}{N}\sum_{j=1}^N e^{-\iu k q_j}\right].
\end{equation}
Thus
\begin{equation}
    F_{p_i}\big(\bm{x}(t'),t'\big)
    =
    -\iu \int_k e^{\iu k q_i(t')} k \, v(k,t') \left[\frac{1}{N}\sum_{j=1}^N e^{-\iu k q_j(t')}\right].
\end{equation}
Then
\begin{equation}
    F_{p_i}\left(\frac{\delta}{\iu \delta \bm{J}(t')},t'\right) 
    =
    -\iu \int_k \hat{\rho}_i(-k,t') k \, v(k,t') \frac{\hat{\rho}(k,t')}{N}.
\end{equation}
Thus, if we define
\begin{equation}
    B(k,t') := \sum_{i=1}^N k\cdot \frac{\delta}{\delta K_{p_i}(t')}\hat{\rho}_i(-k,t'),
\end{equation}
then
\begin{align}
    \mathbb{E}[\tilde\rho_1(k_1,t_1)\cdots \tilde\rho_n(k_n,t_n)]
    = \left. \hat \rho_1(k_1,t_1)\cdots \hat \rho_n(k_n,t_n) \exp\left[-\iu \int_0^T \diff t'\int_k \hat B(k,t') v(k,t') \frac{\hat{\rho}(k,t')}{{N}} \right] Z[\bm{J},\bm{K}]\right|_{\bm{J},\bm{K}=0} .
\end{align}
Now expand the exponential. At zeroth order,
\begin{align}
    \mathbb{E}[\tilde\rho_1(k_1,t_1)\cdots \tilde\rho_n(k_n,t_n)]
    & = \hat \rho_1(k_1,t_1)\cdots \hat \rho_n(k_n,t_n)Z[\bm{J},\bm{K}]\Big|_{\bm{J},\bm{K}=0}, \\
    & =
    \hat \rho_1(k_1,t_1)\cdots \hat \rho_n(k_n,t_n)Z[\bm{J}+\bm{L}_0,\bm{K}]\Big|_{\bm{J},\bm{K}=0} \\
    & = Z[\bm{L}_0,0],
\end{align}
where we have defined
\begin{equation}
    \bm{L}_0(u) = 
    -
    \sum_{i=1}^n 
    \delta_{\rm D}(u-t_i)
    \begin{bmatrix}
    k_i \\ 0
    \end{bmatrix}
    \otimes
    e_i.
\end{equation}
At first order,
\begin{align}
    \mathbb{E}[\tilde\rho_1(k_1,t_1)\cdots \tilde\rho_n(k_n,t_n)] = Z[\bm{L}_0,0] -\iu
    \hat \rho_1(k_1,t_1)\cdots \hat \rho_n(k_n,t_n) \left.\int_0^T \diff t'\int_k \hat B(k,t') v(k,t') \frac{\hat{\rho}(k,t')}{{N}} Z[\bm{J},\bm{K}]\right|_{\bm{J},\bm{K}=0}.
\end{align}
Moving all density operators to the right,
\begin{align}
    & \mathbb{E}[\tilde\rho_1(k_1,t_1)\cdots \tilde\rho_n(k_n,t_n)] \notag \\
    & = Z[\bm{L}_0,0] -\iu\left. \sum_{i=1}^N \int_0^T \diff t'\int_k v(k,t')
    k\cdot \frac{\delta}{\delta K_{p_i}(t')}\hat \rho_1(k_1,t_1)\cdots \hat \rho_n(k_n,t_n) \hat{\rho}_i(-k,t')\frac{\hat{\rho}(k,t')}{{N}} Z[\bm{J},\bm{K}]\right|_{\bm{J},\bm{K}=0} .
\end{align}
Now we invoke the exchangeability assumption \eqref{e:exchangeable}, which justifies the following replacement,
\begin{equation}
    \hat{\rho}(k,t') \approx N\hat{\rho}_{n+1}(k,t').
\end{equation}
Then at first order,
\begin{align}
    & \mathbb{E}[\tilde\rho_1(k_1,t_1)\cdots \tilde\rho_n(k_n,t_n)] \notag \\
    & \approx Z[\bm{L}_0,0] -\iu \left. \sum_{i=1}^{N} \int_0^T \diff t'\int_k v(k,t')
    k\cdot \frac{\delta}{\delta K_{p_i}(t')}
    \hat \rho_1(k_1,t_1)\cdots \hat \rho_n(k_n,t_n) \hat \rho_i(-k,t')  \frac{N\hat\rho_{n+1}(k,t')}{{N}}Z[\bm{J},\bm{K}]\right|_{\bm{J},\bm{K}=0}, \\
    & = Z[\bm{L}_0,0] -\iu \left. \sum_{i=1}^{N} \int_0^T \diff t'\int_k v(k,t')
    k\cdot \frac{\delta}{\delta K_{p_i}(t')}
    \hat \rho_1(k_1,t_1)\cdots \hat \rho_n(k_n,t_n) \hat\rho_{n+1}(k,t') \hat \rho_i(-k,t') Z[\bm{J},\bm{K}]\right|_{\bm{J},\bm{K}=0}.
\end{align}
Evaluating the functional derivatives with respect to $\bm{J}$ and setting $\bm{J}=0$ gives
\begin{align}
    \mathbb{E}[\tilde\rho_1(k_1,t_1)\cdots \tilde\rho_n(k_n,t_n)]
    & = Z[\bm{L}_0,0] -\iu \left. \sum_{i=1}^{N} \int_0^T\diff t'\int_k v(k,t')
    k\cdot \frac{\delta}{\delta K_{p_i}(t')}
    Z[\bm{L}_0 + \bm{L}_i,\bm{K}]\right|_{\bm{K}=0},
\end{align}
where we have defined
\begin{equation}
    \bm{L}_i(u) = 
    -
    \delta_{\rm D}(u-t')
    \begin{bmatrix}
    k \\ 0
    \end{bmatrix}
    \otimes
    e_{n+1}
    -
    \delta_{\rm D}(u-t')
    \begin{bmatrix}
    -k \\ 0
    \end{bmatrix}
    \otimes
    e_i,
\end{equation}
and we have left the dependence of $\bm{L}_i$ on $t'$ and $k$ implicit. Now consider the functional derivatives with respect to $\bm{K}$,
\begin{align}
    \left.\frac{\delta}{\delta K_{p_i}(t')}
    Z[\bm{L}_0 + \bm{L}_i,\bm{K}]\right|_{\bm{K}=0}
    & = 
    \left. \iu Z[\bm{L}_0 + \bm{L}_i,0]\frac{\delta}{\delta K_{p_i}(t')}\int_0^T\diff u \big(\bm{L}_0(u) + \bm{L}_i(u)\big) \cdot \bar{\bm{x}}[\bm{K}](u)\right|_{\bm{K}=0}.
\end{align}
Recalling \eqref{e:xbar}, \eqref{e:L0}, \eqref{e:Li} and \eqref{e:Green},
\begin{align}
    \bm{L}_0(u)
    & = 
    -
    \sum_{j=1}^n 
    \delta_{\rm D}(u-t_j)
    \begin{bmatrix}
    k_j \\ 0
    \end{bmatrix}
    \otimes
    e_j, \\
    \bm{L}_i(u) & := 
    -
    \delta_{\rm D}(u-t')
    \begin{bmatrix}
    k \\ 0
    \end{bmatrix}
    \otimes
    e_{n+1}
    -
    \delta_{\rm D}(u-t')
    \begin{bmatrix}
    -k \\ 0
    \end{bmatrix}
    \otimes
    e_i, \\
    \bar{\bm{x}}[\bm{K}](u)
    & = \bm{G}(u,0) \bm{x} + \int_0^u \diff \tau \, \bm{G}(u,\tau) \bm{K}(\tau), \\
    & =\sum_{j=1}^N \left[G(u,0)x_j + \int_0^u \diff \tau \, G(u,\tau) K_j(\tau)\right] \otimes e_j.
\end{align}
In taking the inner product between $\bm{L}_0(u)$ and $\bar{\bm{x}}[\bm{K}](u)$, notice that only the first $n$ terms survive,
\begin{align}
    \int_0^T\diff u \, \bm{L}_0(u) \cdot \bar{\bm{x}}[\bm{K}](u)
    & = 
    - \sum_{j=1}^n
    \int_0^T\diff u \, \delta_{\rm D}(u-t_j) 
    \begin{bmatrix}
    k_j \\ 0
    \end{bmatrix}
    \cdot
    \left[G(u,0)x_j + \int_0^u \diff \tau \, G(u,\tau) K_j(\tau)\right].
\end{align}
Recalling that $T > t_j$ for all $j\in[n]$, the integral can be evaluated to give
\begin{align}
    \int_0^T\diff u \, \bm{L}_0(u) \cdot \bar{\bm{x}}[\bm{K}](u)
    & =
    -\sum_{j=1}^n 
    \begin{bmatrix}
    k_j \\ 0
    \end{bmatrix} \cdot  \left[G(t_j,0)x_j + \int_0^{t_j} \diff \tau \, G(t_j,\tau) K_j(\tau)\right], \\
    & =
    -\sum_{j=1}^n 
    \begin{bmatrix}
    k_j \\ 0
    \end{bmatrix} \cdot  \left(G(t_j,0)x_j + \int_0^{t_j} \diff \tau \, G(t_j,\tau)
    \begin{bmatrix}
        K_{q_j}(\tau) \\
        K_{p_j}(\tau)
    \end{bmatrix}
    \right).
\end{align}
Then
\begin{align}
    k\cdot\frac{\delta}{\delta K_{p_i}(t')}
    \int_0^T\diff u \, \bm{L}_0(u) \cdot \bar{\bm{x}}[\bm{K}](u)
    & =
    -\sum_{j=1}^n 
    \begin{bmatrix}
    k_j \\ 0
    \end{bmatrix} \cdot \int_0^{t_j} \diff \tau \, G(t_j,\tau)
    \begin{bmatrix}
        0 \\
        \delta_{ij} \delta_\textrm{D}(\tau-t') k
    \end{bmatrix}, \\
    & =
    \begin{cases}
    -
    \begin{bmatrix}
    k_i \\ 0
    \end{bmatrix} 
    \cdot
    G(t_i,t') 
    \begin{bmatrix}
    0 \\
    k
    \end{bmatrix} 
    \theta(t_i - t'),
    & i \in [n] \\
    0, & i>n
    \end{cases}
    .
\end{align}
Now,
\begin{align}
    \int_0^T\diff u \, \bm{L}_i(u) \cdot \bar{\bm{x}}[\bm{K}](u)
    & = 
    -\int_0^T\diff u \, \delta_\textrm{D}(u-t')
    \begin{bmatrix}
        k \\
        0
    \end{bmatrix}
    \cdot
    \left[G(u,0)x_{n+1} + \int_0^u \diff \tau \, G(u,\tau) K_{n+1}(\tau)\right] \notag \\
    & \mathrel{\phantom{=}}
    -\int_0^T\diff u \, \delta_\textrm{D}(u-t')
    \begin{bmatrix}
        -k \\
        0
    \end{bmatrix}
    \cdot
    \left[G(u,0)x_i + \int_0^u \diff \tau \, G(u,\tau) K_i(\tau)\right], \\
    & =
    -
    \begin{bmatrix}
    k \\ 0
    \end{bmatrix} \cdot  \bigg[G(t',0)x_{n+1} + \int_0^{t'} \diff \tau \, G(t',\tau) K_{n+1}(\tau)\bigg] \notag \\
    & \mathrel{\phantom{=}}-
    \begin{bmatrix}
    -k \\ 0
    \end{bmatrix} \cdot  \bigg[G(t',0)x_{i} + \int_0^{t'} \diff \tau \, G(t',\tau) K_i(\tau)\bigg].
\end{align}
Then,
\begin{align}
    k\cdot\frac{\delta}{\delta K_{p_i}(t')}
    \int_0^T\diff u \, \bm{L}_i(u) \cdot \bar{\bm{x}}[\bm{K}](u)
    & =
    -
    \begin{bmatrix}
    k \\ 0
    \end{bmatrix} \cdot \int_0^{t'} \diff \tau \, G(t',\tau)
    \begin{bmatrix}
        0 \\
        \delta_{i,n+1} \delta_\textrm{D}(\tau-t') k
    \end{bmatrix} \notag\\
    & \mathrel{\phantom{=}}-
    \begin{bmatrix}
    -k \\ 0
    \end{bmatrix} \cdot \int_0^{t'} \diff \tau \, G(t',\tau)
    \begin{bmatrix}
        0 \\
        \delta_\textrm{D}(\tau-t') k
    \end{bmatrix}, \\
    & = 0.
\end{align}
Thus,
\begin{align}
    k\cdot \frac{\delta}{\delta K_{p_i}(t')}\int_0^T\diff u \big(\bm{L}_0(u) + \bm{L}_i(u)\big) \cdot \bar{\bm{x}}[\bm{K}](u) & =
    \begin{cases}
    -
    \begin{bmatrix}
    k_i \\ 0
    \end{bmatrix} 
    \cdot
    G(t_i,t') 
    \begin{bmatrix}
    0 \\
    k
    \end{bmatrix} 
    \theta(t_i - t'),
    & i \in \{1, \ldots, n \} \\
    0, & i>n
    \end{cases}.
\end{align}
Plugging back in gives the first-order perturbation result gives
\begin{equation}
    \mathbb{E}[\tilde\rho_1(k_1,t_1)\cdots \tilde\rho_n(k_n,t_n)] = Z[\bm{L}_0,0] - \sum_{i=1}^{n} \int_0^{t_i}\diff t'\int_{k'} v(k',t')
    Z[\bm{L}_0+\bm{L}_i,0] \begin{bmatrix}
    k_i \\ 0
    \end{bmatrix} 
    \cdot
    G(t_i,t') 
    \begin{bmatrix}
    0 \\
    k'
    \end{bmatrix}.
\end{equation}

\bibliographystyle{JHEP}
\bibliography{biblio.bib}

\end{document}